\documentclass[twocolumn,prl]{revtex4}

\usepackage{graphicx}
\usepackage{dcolumn}
\usepackage{amsmath}

\begin{document}
\preprint{HEP/123-qed}

\title[Short Title]{Stability of 1-D Excitons in Carbon
Nanotubes under High Laser Excitations}

\author{G. N. Ostojic}
\author{S. Zaric}
\author{J. Kono}
 \thanks{To whom correspondence should be addressed}
 \homepage{http://www.ece.rice.edu/~kono}
 \email{kono@rice.edu}
\affiliation{Department of Electrical and Computer Engineering,
Rice Quantum Institute,
and Center for Nanoscale Science and Technology,
Rice University, Houston, Texas 77005}

\author{V. C. Moore}
\author{R. H. Hauge}
\author{R. E. Smalley}
\affiliation{Department of Chemistry,
Rice Quantum Institute, and Center for Nanoscale Science and
Technology, Rice University, Houston, Texas 77005}

\date{\today}

\begin{abstract}

Through ultrafast pump-probe spectroscopy with intense pump
pulses and a wide continuum probe, we show that interband
exciton peaks in single-walled carbon nanotubes (SWNTs) are
extremely stable under high laser excitations. Estimates
of the initial densities of excitons from the excitation
conditions, combined with recent theoretical calculations of
exciton Bohr radii for SWNTs, suggest that their positions do
not change at all even near the Mott density.  In addition, we
found that the presence of lowest-subband excitons broadens all
absorption peaks, including those in the second-subband range,
which provides a consistent explanation for the complex spectral
dependence of pump-probe signals reported for SWNTs.

\end{abstract}

\maketitle

Optically-excited electron-hole ($e$-$h$) pairs in a
semiconductor provide a rich system for the study of carrier
interaction effects.  Depending on the density, they exhibit
qualitatively different spectral features of bound
and unbound carriers. Excitons (or bound $e$-$h$ pairs) are
stable when the Bohr radius is much
smaller than the inter-exciton distance.  As the former
approaches the latter,
the Mott transition \cite{Mott61PM} occurs, transforming the
insulating excitonic gas into a metallic $e$-$h$ plasma.  This
scenario is complicated in real systems by other
interaction effects such as band gap renormalization (BGR)
and biexcitonic correlations.  In addition, optical gain
develops and coherent processes can dominate the emission
spectra.  There have been a number of studies on highly-excited
3-D and 2-D semiconductor systems
\cite{Haken75Book,HaugKoch04Book}.

1-D excitons are expected to be different: the exciton
binding energy in an ideal 1-D system is infinite
\cite{Loudon59AJP}, the 1-D Sommerfeld factor is less than 1
\cite{OgawaTakagahara91both}, and radiative lifetimes are
intrinsically longer \cite{CitrinetAl92PRL}.
High-density excitons in semiconductor quantum wires
\cite{KaponetAl89PRL,WegscheideretAl93PRL,AmbigapathyetAl97PRL,AkiyamaetAl02SSC,AkiyamaetAl03PRB}
have exhibited a range of novel (and conflicting) results
regarding gain, BGR, and Mott transitions.  The magnitude of 1-D
BGR is still under debate, but it appears to be a common
observation that 1-D excitons are stable up to very high
densities
\cite{WegscheideretAl93PRL,AmbigapathyetAl97PRL,AkiyamaetAl02SSC}.
This stability, while good for device applications, is
not fully understood, and the density at which gain should
appear is a subject of controversy \cite{1d-gain}.

Excitons in carbon nanotubes can provide new insight into these
long-standing issues.  With much smaller diameters (and thus
larger subband separations and binding energies),
these excitons should allow one to study high-density regimes
without having to take into account the population of
higher subbands.  Since the success of preparing
individually-suspended single-walled carbon nanotubes (SWNTs)
\cite{OconnelletAl02Science,BachiloetAl02Science}, their optical
properties have been intensively studied. While theoretical
understanding of linear optical properties is progressing
\cite{Ando97JPSJ,KaneMele03PRL,SpataruetAl04PRL,ChangetAl04PRL,Pederson04Carbon,PerebeinosetAl04PRL},
their nonlinear optical properties, especially under high
excitations, remain unexplored.  Recent time-resolved studies on
bundled \cite{LauretetAl03PRL,KorovyankoetAl04PRL} and unbundled
\cite{OstojicetAl04PRL,MaetAl04JCP,HagenetAl04APA,WangetAl04PRL,HuangetAl04PRL}
SWNTs have raised an array of questions on the origin of
non-radiative recombination and the values of intrinsic
radiative lifetimes. In particular, strongly
wavelength-dependent non-degenerate pump-probe data
\cite{LauretetAl03PRL,KorovyankoetAl04PRL,MaetAl04JCP} have
produced differing interpretations.

\begin{figure}
\includegraphics [scale=0.46] {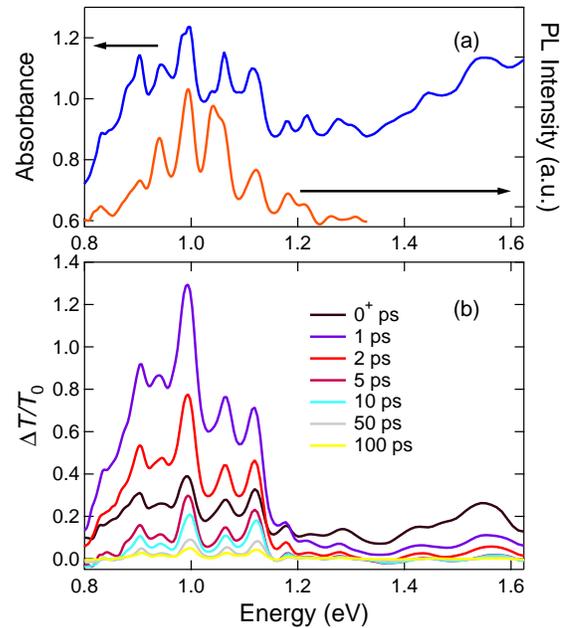}
\caption{(a) Linear absorption (blue line, left axis) and PL
(red line, right axis) spectra.  The PL was taken with CW 1.6 eV
(775 nm) excitation.  (b) Chirp corrected differential
transmission spectra for for various time delays.} \end{figure}

\begin{figure}
\includegraphics [scale=0.48] {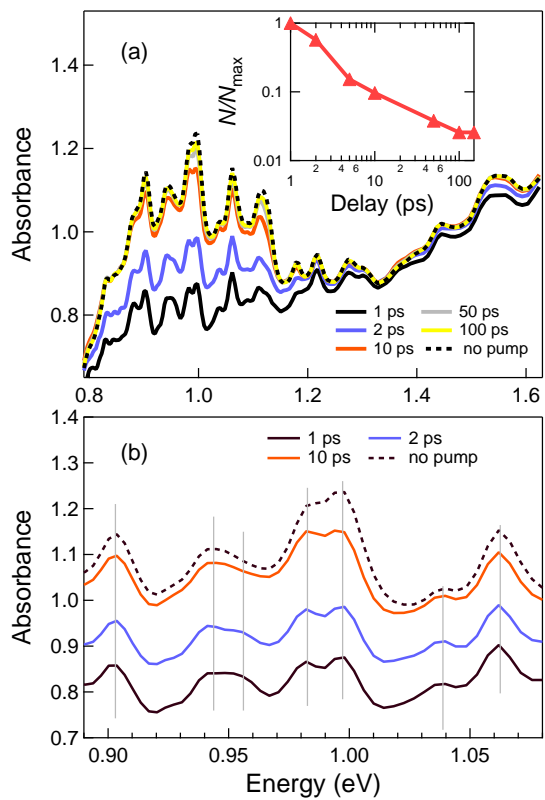}
\caption{(a) Absorbance spectra at different time
delays, reconstructed from the differential transmission data in
Fig.~1.  The inset shows the time decay of the normalized
$e$-$h$ pair density, obtained by spectrally integrating the
absorbance change at each time delay.  (b) Absorbance spectra
within the first subband range at different time delays, showing
that exciton peaks do not shift.} \end{figure}

In this Letter, we report results of non-degenerate pump-probe
spectroscopy with a strong pump beam and a white-light
continuum probe. From excitation conditions we estimated the
initial pair density to be $\sim 5 \times 10^6$ cm$^{-1}$. This,
combined with recent calculations of the Bohr
radius (2-5 nm)
\cite{SpataruetAl04PRL,ChangetAl04PRL,Pederson04Carbon},
indicates that the initial density is comparable to the Mott
density.  However, the band edge absorption peaks were very
stable, their positions showing no variation with time
(i.e., density), similar to the observations in quantum wires
\cite{WegscheideretAl93PRL,AmbigapathyetAl97PRL,AkiyamaetAl02SSC}.
Furthermore, we found that there are long-lived (up to $\sim 100$
ps), pump-probe signals in the second-subband transition range,
even though there are no real carriers left.  Through
spectral analysis we concluded that the absorption peaks are
broadened by the excitons present in the first subbands.  This
result provides a new and consistent scenario for the complex
pump-probe dynamics.


The samples used in our experiments were
micelle-suspended in D$_2$O, which showed a
number of chirality-dependent peaks in photoluminescence (PL)
and absorption \cite{OconnelletAl02Science}.  The first
($E_{11}$) and second ($E_{22}$) subband transition energies
were determined through PL excitation (PLE) spectroscopy. They
possess long pump-probe decay times,
especially when excited resonantly \cite{OstojicetAl04PRL}.
Similar lifetimes have been reported using
time-resolved PL measurements
\cite{MaetAl04JCP,HagenetAl04APA,WangetAl04PRL},
indicating that these long lifetimes are related to radiative
interband recombination.

We used an optical parametric amplifier (OPA) pumped by a
chirped-pulse amplifier (CPA2010, Clark-MXR) that produced
150-fs pulses at 1 kHz. As a pump, either the OPA beam covering
0.07--2.5 eV or the CPA beam (1.6 eV or 775 nm) was used.  Pump
fluences up to $\sim 1$ mJ/cm$^{2}$ were used.  As a probe, we
used a white light continuum generated by focusing a small
portion of the CPA beam onto a sapphire crystal, covering
wavelengths from the visible to the infrared.
Both beams were focused onto a 5-mm thick sample cell with pump
and probe diameters 1 mm and 0.8 mm, respectively.  After
passing through the sample, the probe was spectrally resolved
with a monochromator and detected by an InGaAs photodiode.
Chirp in the probe was measured through a second
harmonic generation technique, and it was taken into account in
all data analysis.  The pulse width of the probe (i.e., the
temporal resolution of the setup) was 270 fs.  To detect the
small change in probe transmission $\Delta T$, we synchronously
chopped the pump beam at 500 Hz, which blocked every other pump
pulses.  The probe pulse train was then sampled in a box car
integrator that electronically inverted every other signals.  A
computer then collected the pulses and separate them to extract
pump influenced ($T$) and non-influenced
transmission ($T_{0}$).


Absorption and PL spectra are shown
in Fig.~1(a).  The PL was excited by CW 775 nm
radiation. Differential transmission spectra with a 775 nm pump
are shown in Fig. 1(b) for different time delays. At 0 ps,
positive $\Delta T$ exists both in the $E_{11}$ and $E_{22}$
ranges. However, during the first 1 ps, the signal in the
$E_{11}$ range rapidly increases at the expense of the signal in
the $E_{22}$ range, indicating a fast intraband (i.e.,
intersubband) carrier decay.  At subsequent time delays, the
overall shape of the spectra remains the same, and the amplitude
decays slowly.

The evolution of absorbance, obtained from differential
transmission, is shown in Fig.~2(a).  Absorption quenching
up to $\sim$ 80\% is seen at 1 ps but it recovers as
time progresses.  At 100 ps, the absorption spectrum coincides
with that taken at negative time delays, indicating that a
majority of carriers have already recombined.  The inset shows
the estimated pair density versus time, normalized to
the value at 1 ps.  The density at each time delay was
deduced from the amount of spectrally-integrated absorption
change.  Figure 2(b) shows an expanded view in the 0.9--1.1 eV
range. It is seen that the {\em absorption peak positions
do not change at all with time, i.e., with density}, while
linewidths increase slightly.


\begin{figure}
\includegraphics [scale=0.45] {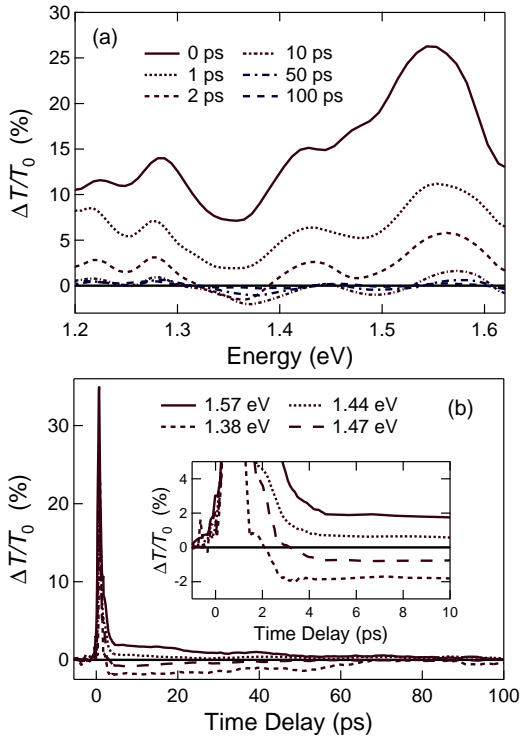}
\caption{(a) Differential transmission spectra at
different time delays in the second subband range.  (b)
Differential transmission dynamics at four different probe photon
energies.  At some probe photon energies the signal becomes
negative (corresponding to absorption increase).}
\end{figure}

We also found that the presence of first subband excitons
modifies the absorption in the second subband.  Figure 3(a)
shows $\Delta T/T_0$ as a function of probe energy, mostly
in the second subband range.  This continuous probing reveals
previously-unobserved {\em oscillatory behavior}. Especially
after 1 ps, when all carriers have decayed into the first
subband, the {\em sign of $\Delta T$ sensitively depends on the
probe energy.} Positive $\Delta T$ (i.e., absorption bleaching)
is located mostly around absorption peaks while negative $\Delta
T$ (i.e., photoinduced absorption) exists in the vicinity of
absorption dips. Examples of $\Delta T/T_0$ versus time delay
for selected probe energies are shown in Fig.~3(b).
Upon close examination, this oscillatory behavior is also
seen in the first subband range,
superimposed onto the positive band filling signal.
The density of photo-excited $e$-$h$ pairs ($n_X$) was estimated
as follows.  For typical 1-nm-diameter nanotubes that
are abundant in the sample [say, (8,7) tubes], the linear mass
density is calculated to be 2.4 $\times$ 10$^{-11}$ mg/cm.
This, combined with the mass density (52.4 mg/l from linear
absorption with Beer's law) and the average tube length (150
nm from atomic force microscopy), provides $5.7 \times
10^{11}$ as the total number of tubes within the pump-excited
volume (3.9 mm$^{3}$).  The number of absorbed photons (and
thus created $e$-$h$ pairs) in the same volume is estimated to
be 2.1 $\times$ 10$^{13}$ from the nonlinear absorbance (1.043
at 0 ps for 1.6 eV) and the pump energy (12 $\mu$J).  These
numbers yield $n_X$ $\approx$ 5 $\times$ 10$^6$ cm$^{-1}$ (or
an average separation of $\sim$ 2 nm).  [Note that the average
length (150 nm) cancels out and does not affect $n_X$.]
A possible source of uncertainty in $n_X$ is the existence of
residual bundles, which can absorb the pump and make the
estimated $n_X$ larger. However, the clear appearance of peaks in
absorption indicates that absorption from unbundled SWNTs is
dominant. A possible error may also come from the neglect of
scattering, but a recent study \cite{VivienetAl02Carbon}
indicates that scattering is of small importance for low
repetition-rate femtosecond pulses. Finally, we did not take
into account any contribution of two-photon absorption, which
should lead to the creation of $e$-$h$ pairs at 3.2 eV, mostly
in metallic nanotubes, and thus reduce the pair density in
semiconductor nanotubes.

Recent theoretical calculations using different methods
\cite{SpataruetAl04PRL,ChangetAl04PRL,Pederson04Carbon}
have reported 2-5 nm for the Bohr radii
($a^*_B$) of excitons in SWNTs with a $\sim$ 1 nm diameter.
This suggests that $r_s = n_X a^*_B \sim$ 1.  Namely, right
after they are created by the pump pulse, the exciton density is
similar to the 1-D Mott density.  Since the photoinduced carrier
density decays to almost zero in $\sim$ 200 ps, this translates
that we are monitoring the behavior of exciton peaks in a widely
varying density range, going from the initial high-density regime
to the final dilute limit. Nonetheless, the observed absorption
peaks are very stable in position.  This stability is
similar to what has been observed in high-quality GaAs quantum
wires
\cite{WegscheideretAl93PRL,AmbigapathyetAl97PRL,AkiyamaetAl02SSC},
implying a unique nature of excitons in 1-D systems.

Previous pump-probe spectroscopy studies on bundled
nanotubes \cite{LauretetAl03PRL,KorovyankoetAl04PRL}
exhibited similar probe-energy-dependent bleaching and
absorption behaviors, but different mechanisms were proposed.
Lauret {\it et al}.~\cite{LauretetAl03PRL} created carriers in
the first subband and observed increased absorption at selected
probe energies, which was interpreted as a redshift of the
``plasmon'' absorption peak. Korovyanko {\it et
al}.~\cite{KorovyankoetAl04PRL} explored various probe energies and concluded that a
global red shift is not a consistent picture to explain their
data; instead they attributed the photoinduced absorption
observed at some probe energies to excitonic intersubband
transitions.  Neither of the proposed mechanisms can explain our
data. Since our sample consists of unbundled SWNTs that exhibit
chirality-dependent peaks, we can readily see that we have no
global red shift or photo-induced absorption peaks corresponding
to intersubband-like transitions.

\begin{figure}
\includegraphics [scale=0.54] {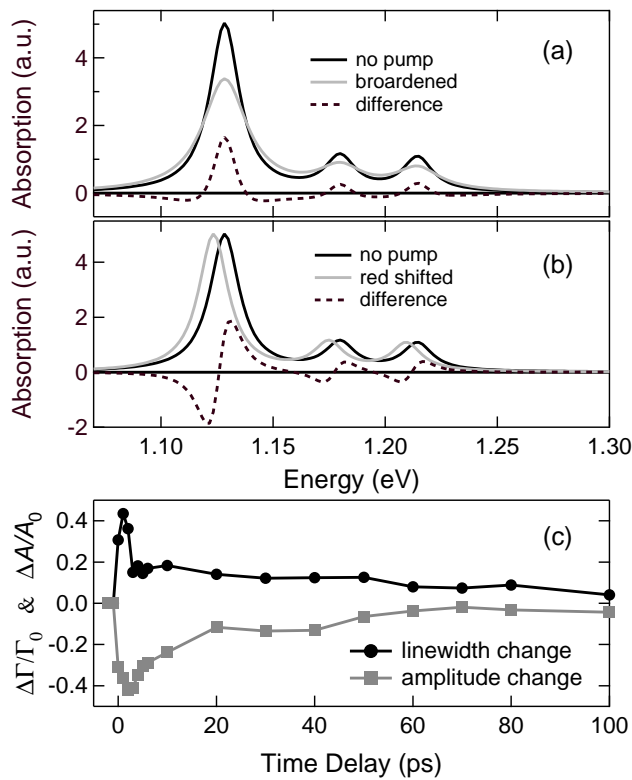}
\caption{Simulated differential absorption spectra for three
closely-lying lines with different amplitudes based on (a)
line broadening and (b) a rigid red shift.  (c) Obtained
best fit Lorentzian parameters for the 1.17 eV absorption peak
of (10,2) nanotubes.} \end{figure}

We believe that the probe-energy-dependent sign of differential
transmission can be consistently explained in terms of
carrier-induced peak broadenings.  Figure 4(a) depicts the idea,
which is a simulation result based on three Lorentzian peaks.
All three tube types are assumed to have a half-width at half
maximum of 8 meV and one of them is assumed to have a larger
amplitude.  It can be seen that the sign of pump-probe signal
should show the observed oscillatory behavior.  It can also be
seen that the change of a small peak can be significantly
influenced by a neighboring large peak. Figure 4(b) is a
simulation result based on a global red shift; this model fails
to explain any of the observed features.

To analyze the photoinduced broadening in more detail, we
concentrated on (10,2) tubes, which have an $E_{11}$ peak
reasonably separated from other tubes and is relatively weakly
populated (i.e., the positive band filling signal is
small).  The evolution of the extracted Lorentzian parameters
(peak amplitude $\Delta A/A_0$ and linewidth $\Delta
\Gamma/\Gamma_0$) normalized to the negative delay values are
shown in Fig.~4(c).
Broadening and amplitude change follow the fast rise seen in
$\Delta T/T_{0}$.  However, the subsequent decay is
slightly faster for the broadening.  The amplitude decay is
related to state filling but also can result from the screening
of the Coulomb interaction. The latter effect produces broadening
although carrier-phonon scattering and carrier-carrier scattering
can have important roles in broadening in the regime of high
carrier density, which could explain the faster decay of
broadening compared to the amplitude change.

In conclusion, we have performed non-degenerate pump-probe
spectroscopy on micelle-suspended SWNTs and found 1)
that interband exciton peaks are very stable even at very high
densities and 2) that second-subband absorption peaks are
broadened by the excitons present in the first subbands, which
provides a new and consistent explanation for the complex
behavior of pump-probe signals reported for bundled and
unbundled SWNTs.






We gratefully acknowledge support from the Robert A.~Welch
Foundation (Grant No.~C-1509), the Texas Advanced Technology
Program (Project No.~003604-0001-2001), and the National Science
Foundation (Grant Nos.~DMR-0134058 and DMR-0325474).
 

\end{document}